\documentstyle[times,pramana,epsf,floats]{ias}
\begin{document}
\title{Message transfer in a communication network}

\author{Satyam Mukherjee and Neelima Gupte}
\address{Department of Physics, Indian Institute of Technology Madras, Chennai - 600036}
\abstract{We study message transfer in a  $2-d$ communication network of
regular nodes and randomly distributed hubs. We study both single
message transfer and multiple message transfer on the lattice. 
The average travel time for single messages travelling between source
and target pairs of fixed separations shows $q-$exponential behaviour as
a function of hub density with a characteristic power-law tail,
indicating a rapid drop in the average travel time as a function of hub
density. This
power-law tail arises as a consequence of the log-normal distribution of 
travel times seen at high hub densities. When many messages travel on
the lattice, a congestion-decongestion transition can be seen. 
The waiting times of messages  
in the congested phase show a Gaussian  distribution,  whereas  the
decongested phase shows a log-normal distribution. Thus, the congested
or decongested behaviour is encrypted in the behaviour of the waiting
time distributions.}

\maketitle
\section{Introduction}
Transport processes on networks have  been a topic of intensive research in recent years. Examples of transport processes on networks include the traffic of information packets\cite{tadic,thurn,guilera,rosvall}, transport processes on biological 
networks \cite{a,c}, and road traffic. The structure and topology of the
network, as well as the mechanism of transport, have been seen to play
crucial roles in the optimisation of the efficiency of the transport
process\cite{kim}. It is therefore important to study this interplay in
the context of realistic networks so that their performance can be optimised.

The gradient mechanism of transport has been shown to be very efficient
for message communication. The topology of the substrate network on
which the gradient is set up, has
proved to be extremely important in the context of the jamming and
congestion of messages. The gradient congests very rapidly when based on
a random network substrate, but is much less prone to
congestion on a scale-free substrate. Here, we study the gradient
mechanism on a two dimensional network of nodes and hubs which
incorporates local clustering and geographic separations. 
We study single message transport as well as multiple message transport 
for this system. In the case of single message transfer, we study the
dependence  average travel times on the hub density, and find that
travel times fall off as a $q-$exponential with a power-law tail  at
high hub densities. The  distribution of travel times is used to analyse
this behaviour. In the case of multiple message transfer, the waiting
time 
distribution of the messages has its characteristic signature, being
normal  in the
congested phase and log-normal in the decongested phase.


\section{Model of Communication Network}
\begin {figure}
\epsfxsize=7.0cm
\centerline{\epsfbox{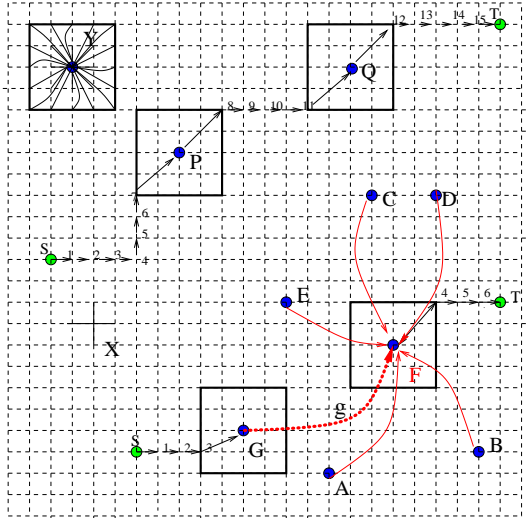}}
\caption{\label{fig:model}(Colour online) A regular two dimensional lattice. {\it X} is an ordinary node with nearest neighbour connections. Each hub has a square influence region (as shown for the hub {\it Y}). A typical path from the source $S$ to the target $T$ is shown with labelled sites. The path S-1-2-3-$\cdots$-7-P-8-$\cdots$-11-Q-12-$\cdots$-T passes through the hubs {\it P} and {\it Q}. Hubs $A - G$ are distributed randomly in the lattice and and each hub is assigned with some message capacity between 1 and 10. In the figure $F$ has maximum capacity 10. The hubs are connected by the gradient mechanism as shown by one way arrows. After the implementation of the gradient mechanism the distance between {\it G} and{\it F} is covered in one step. The gradient path is given by $S-1-2-3-$G$-$g$-$F$-4-5-6-T$.}
\end{figure}

The substrate model on which message communication takes place is shown in Fig.~\ref{fig:model}(a). This is  a regular 2-dimensional lattice  with two types of nodes, the regular nodes, connected to their nearest neighbours (e.g. node $X$ in Fig.~\ref{fig:model}), and hubs at randomly selected locations which are connected to all nodes in their area of influence, a square of side $a$ (e.g. node $Y$ in the same figure). 

To set up the gradient mechanism, we need to assign a  capacity to each
hub, the
hub capacity being defined to be the number of
messages the hub can process simultaneously.
Here, each hub is randomly assigned some message capacity
between one and $C_{max}$.
A gradient flow is assigned from each hub to all the hubs with the
maximum
capacity ($C_{max}$).
Thus, the hubs with lower
capacities are connected to the hubs with highest capacity $C_{max}$ by
the gradient mechanism. See Fig. \ref{fig:model} \cite{mg}.

Two nodes with co-ordinates $(is,js)$ and $(it,jt)$ separated by a fixed distance $D_{st}$ = $|is-it|$ + $|js-jt|$ are  chosen from a lattice of a given size $L^{2}$, and assigned  to be the source and target. We set free boundary conditions. If a message is routed from a source $S$ to a target $T$ on this lattice through the baseline mechanism, it takes the path S-1-2-3-$\cdots$-7-P-8-$\cdots$-11-Q-12-$\cdots$-T as in  Fig.~\ref{fig:model}. After the implementation of the gradient mechanism, the distance between $A$ and $B$ is covered in one step as shown by the link $g$ and a  message is routed along the path $S-1-2-3-$G$-$g$-$F$-4-5-6-T$.
 
\section{Mechanism of single message transfer}

Consider a message that starts from the source $S$ and travels towards a
target $T$. Any node which holds the message at a given time, transfers
the message to the node nearest to it, in the direction which minimises
the distance between the current message holder and the target. If a
constituent node is the current message holder, it sends the message
directly to its own hub. When the hub becomes the current message
holder, the message is sent to the constituent node within the square
region, the choice of the constituent node being made by minimising the
distance to the target. When a hub in the lattice becomes the current
message holder, the message is transferred to the hub connected to the
current message holder by the gradient mechanism, if the new hub is in
the direction of the target, otherwise it is transferred to the
constituent nodes of the current hub. The constituent node is chosen
such that the distance from the target is minimised.
\begin {figure}
\begin{tabular}{cc}
\epsfxsize=6.0cm
{\epsfbox{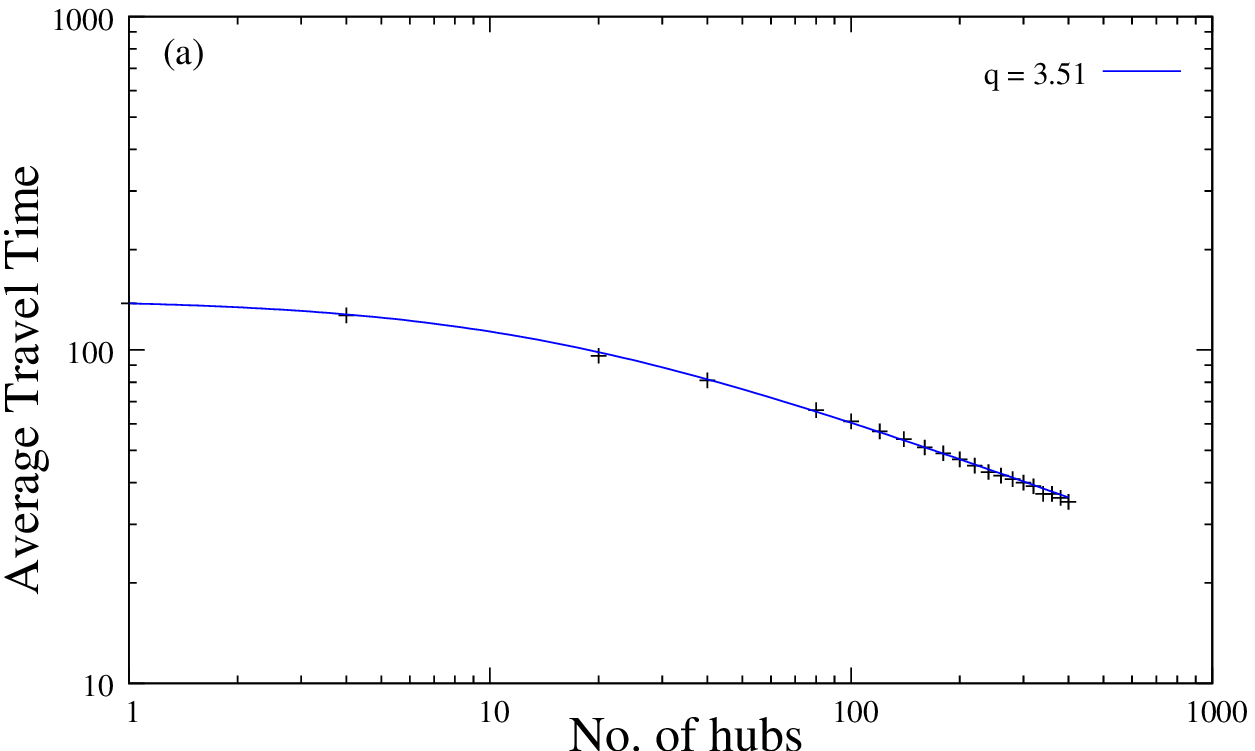}}&
\hspace{0.5cm}
\epsfxsize=6.0cm
{\epsfbox{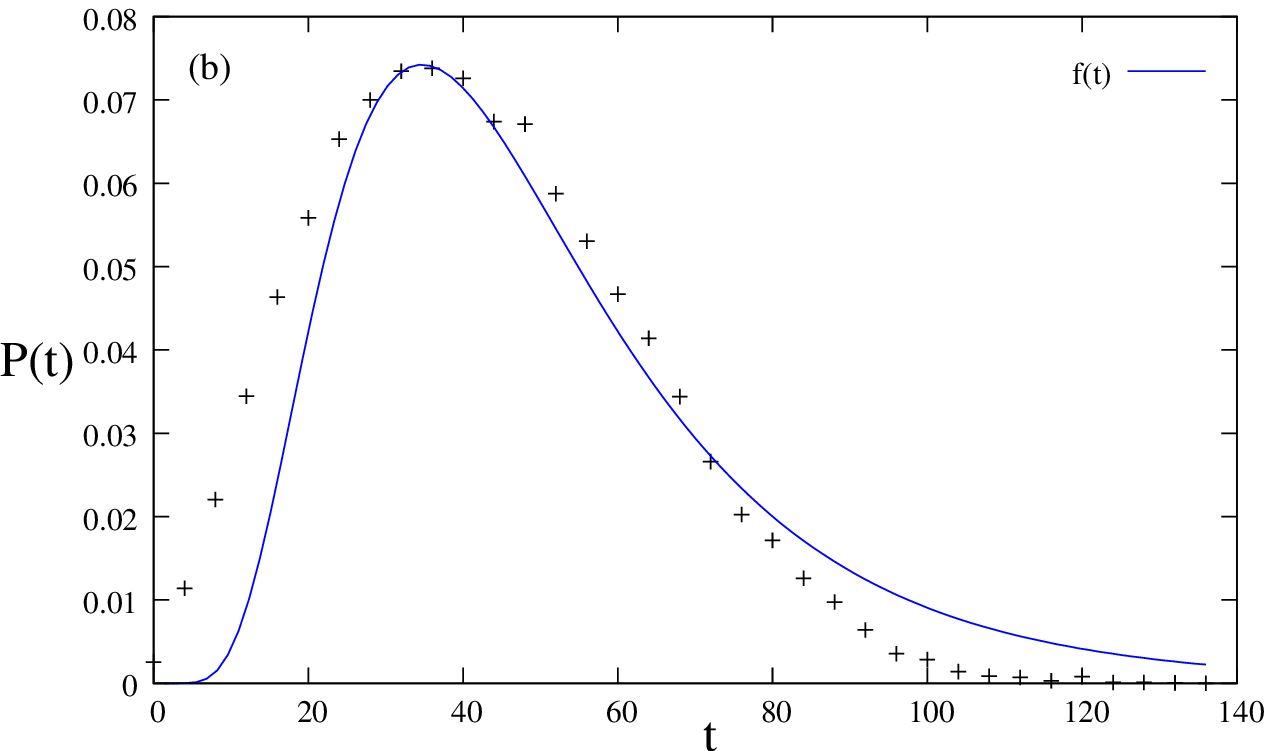}}\\
\end{tabular}
\caption{\label{fig:traveltime}(Colour online)(a) The average travel time as a function of hub density follows a $q$-exponential behavior if the hubs are connected by the gradient mechanism. Here $q$ = 3.51 for the gradient mechanism. (b) The scaled travel time distribution for the gradient mechanism. Different symbols represents lattice of different sizes. The data is fitted by a log-normal distribution (Eq.(1(b))) where ${\mu}=8.78$, ${\sigma}=1.19$.}
\end{figure}

\begin {figure}
\epsfxsize=6.0cm
\centerline{\epsfbox{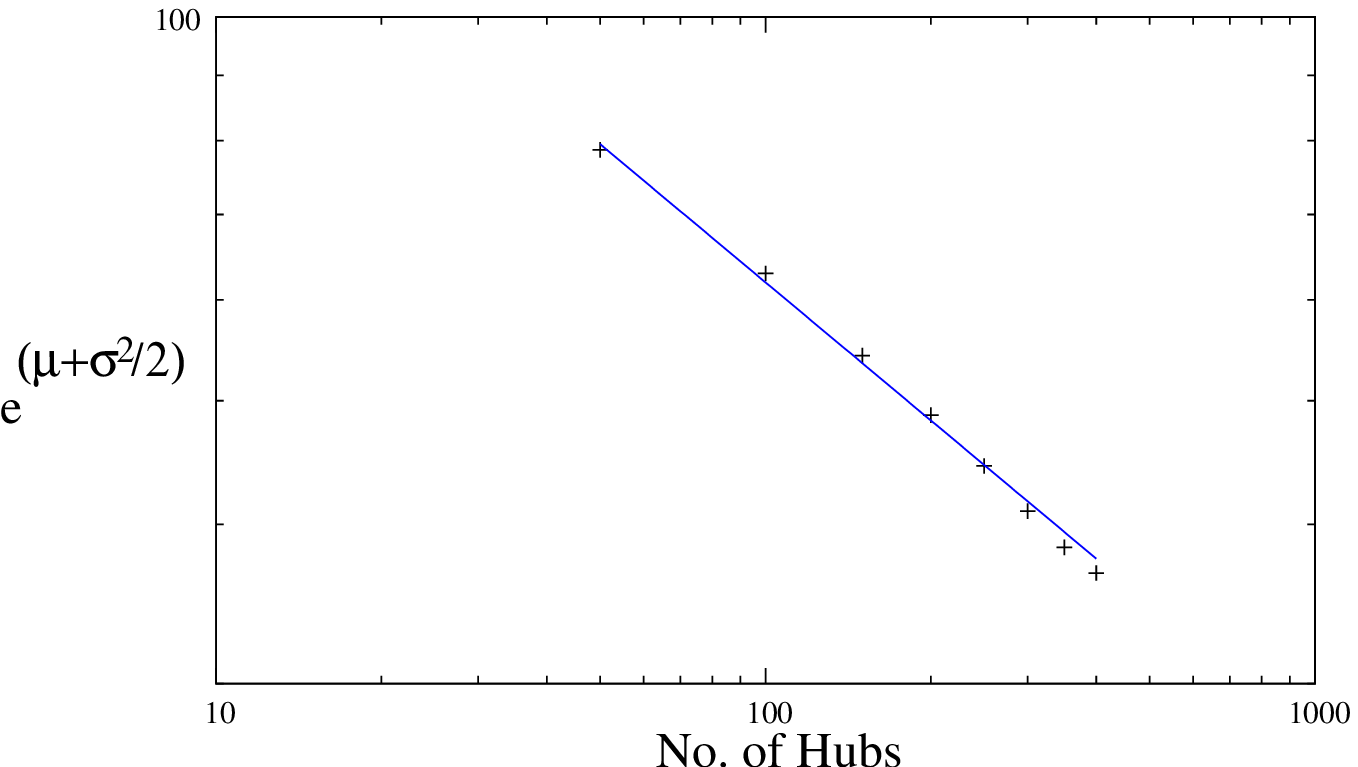}}
\caption{\label{fig:newfig1}(Colour online) Plot of exp($\mu$+$\sigma^{2}$/2) as function of N shows a power law behaviour ($f(x)=Bx^{-\beta}$) for the gradient mechanisms. Here $\beta$ = 0.36 $\pm$ 0.008 for $100\times100$ lattice and $D_{st}$ = 142. }
\end{figure}

The average travel time for a message for a fixed source-target distance, is a good measure of the efficiency of the network. It was observed earlier \cite{mg} that if the hubs in the lattice are connected by the gradient mechanism, the average travel time as a function of hub density shows a $q$-exponential behaviour (Fig.~\ref{fig:traveltime}(a)) of the form $f(x)=A(1-(1-q)\frac{x}{x_{0}})^{\frac{1}{(1-q)}}$ with the parameter $q=3.51$, with a power law tail at higher hub densities. 

The power-law tail at high densities can be understood by looking at the 
travel time distribution at high densities. The travel time distribution
at a hub density of $0.5\%$ is plotted in Fig.~\ref{fig:traveltime}(b) for
lattice size $L$ = 100. 
The travel time distribution shows log-normal behaviour \cite{mg} of the form
\myeqn{f(t)=\frac{1}{t{\sigma}{\sqrt {2\pi}}}exp(-\frac{({\ln}t-{\mu})^{2}}{2{\sigma}^{2}})}{1}
where $\mu$ and $\sigma$ are the mean and standard deviation of the
log-normal fits to the travel time distribution. The mean travel time
for this log-normal distribution $<t>$ is given by
exp($\mu$+$\sigma^{2}$/2)\cite{sornette}.  
We plot this quantity, i.e., exp($\mu$+$\sigma^{2}$/2) as
function of $N$, where $N$ is the number of hubs  in
Fig.~\ref{fig:newfig1}. It is clear that the plot of 
exp($\mu$+$\sigma^{2}$/2)  as a function of N can be fitted by a
power-law  $f(x)=Bx^{-\beta}$ at higher hub densities with the exponent $\beta$ = 0.36 $\pm$ 0.008. 

Thus the average travel time at high hub density is also expected to
show power-law behaviour with a similar exponent. This explains 
the power law tail of the average travel time for single message
transfer plotted in  Fig.~\ref{fig:traveltime}(a) where the exponent
$\beta$ = 0.36 $\pm$ 0.006 \cite{mg} is seen. Finite size scaling was also observed at high hub density for average travel time as a function of hub density and travel time distributions \cite{mg}. The scaled behaviour of travel time distribution fits into a log normal distribution (Eq.1)\cite{mg} where ${\mu}=-1.44$, ${\sigma}=1.47$.

\section{Congestion and Decongestion}

In this section we consider a large number of messages which are created
at the same time and  travel towards their destinations simultaneously.
The hubs on the lattice, and the manner in which they are connected, are
the crucial elements which control the efficiency of message transfer. On the one hand, it is clear that the hubs provide short paths through the lattice. On the other hand, when many messages travel simultaneously on the network, the finite capacity of the hubs can lead to the  trapping of messages in their neighbourhoods, and a consequent congestion or jamming of the network. A crucial quantity which identifies these hubs is called
the coefficient of betweenness centrality (CBC)\cite{braj1}, defined to be the ratio of the number of messages $N_k$ which pass through a given hub $k$ to the total number of messages which run simultaneously  i.e. $CBC=\frac{N_k}{N}$. Hubs with higher CBC are more prone to congestion. 
Here we set up a gradient by identifying the top 5 hubs ranked by their
CBC values and connecting them by the gradient mechanism\cite{mg}. Here,
the capacities of the top five hubs are enhanced proportional to their
CBC values and a gradient is set up between them.

We choose $N$  source-target pairs randomly, separated by a fixed distance $D_{st}$ on the lattice. All sources send messages simultaneously to their respective targets at an initial time $t=0$. The messages are transmitted by a routing mechanism similar to that for single messages, except when the next node or hub on the route is occupied. We carry out parallel updates of nodes.

If the would be recipient node is occupied, then the message waits for a unit time step at the current message holder. If the desired node is still occupied after the waiting time is over, the current node selects any unoccupied node from its remaining neighbors and hands over the message. If all the neighboring nodes are occupied, the message waits at the current node until one of them is free. If the current message holder is the constituent node of a hub which is occupied, the message waits at the constituent node until the hub is free.

We plot the waiting time distributions for the various decongestion schemes mentioned above for lower as well as higher hub densities. We calculate the total waiting time for all the nodes where the messages wait to be delivered to adjacent node along the path of their journey to respective targets. It was observed earlier that for low hub density messages get trapped \cite{mg} at some hubs which lead to congestion in the network. The network is decongested for higher hub densities\cite{mg,braj1}.

\begin {figure}
\begin{tabular}{cc}
\epsfxsize=6.0cm
{\epsfbox{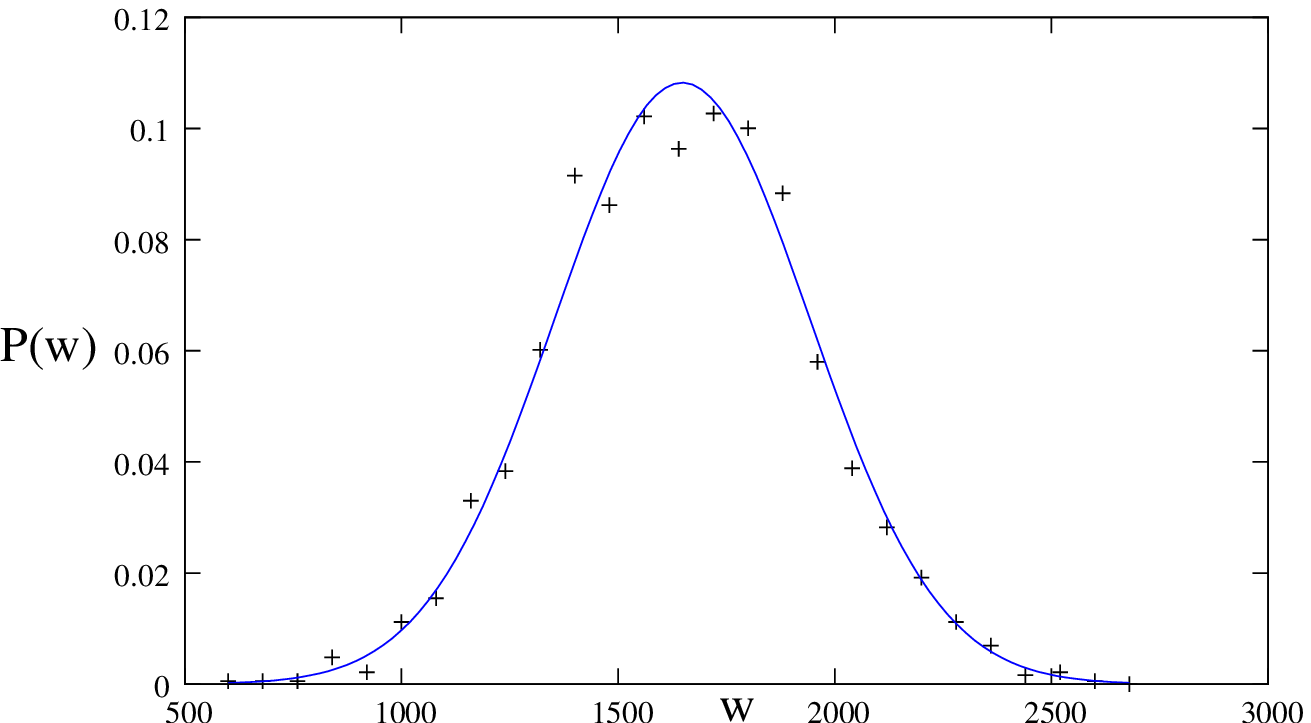}}&
\hspace{0.5cm}
\epsfxsize=6.0cm
{\epsfbox{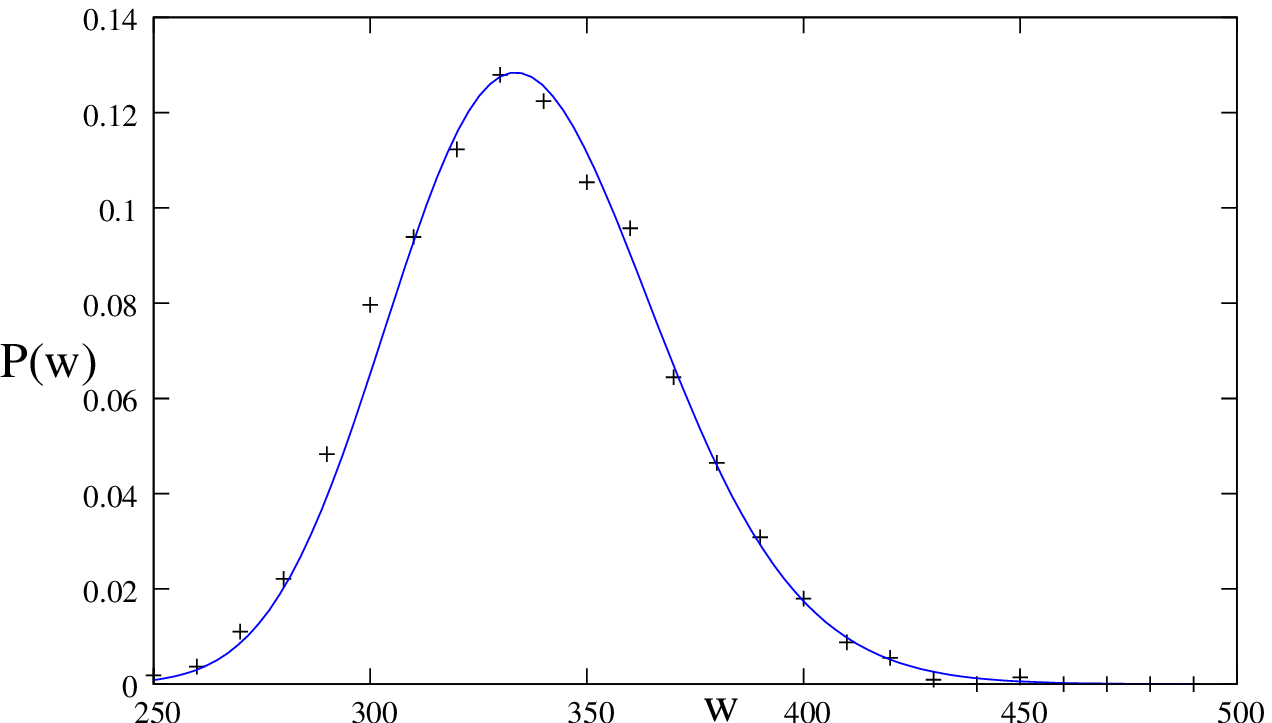}}\\
\end{tabular}
\caption{\label{fig:wait}(Colour online)Waiting time distribution for 2000 messages running simultaneously in the lattice for (a)50 hubs and run time of $30D_{st}$ and (b) 400 hubs and run time of $4D_{st}$ when various decongestion strategies are applied. In (a) the waiting time distribution for messages fits into a Gaussian of the form $\frac{1}{{\sigma}{\sqrt {2\pi}}}exp(-\frac{(w-a)^{2}}{2{\sigma}^{2}})$. The system is congested due to formation of transport traps. The standard deviation $\sigma$ for the gradient mechanism is $294.88$. In (b) all the messages reach their respective destinations. In this case the waiting time distribution fits into a log-normal behaviour of the form $\frac{1}{w{\sigma}{\sqrt {2\pi}}}exp(-\frac{({\ln}w-{\mu})^{2}}{2{\sigma}^{2}})$. The standard deviation $\sigma$ the gradient mechanism is $0.091$. The averaging is done over 200 hub configurations.}
\end{figure}

We allow 2000 messages to run simultaneously in a $100\times 100$ lattice and $D_{st}$ = 142, for 50 hubs and run time of $30d_{st}$. During this phase the messages remain undelivered in the lattice and the system is in the congested regime. The waiting time distribution fits into a Gaussian of the form 
\myeqn{f(w)=\frac{1}{{\sigma}{\sqrt {2\pi}}}exp(-\frac{(w-a)^{2}}{2{\sigma}^{2}})}{2(a)}
\eqnreset
For 400 hubs and $4d_{st}$ all the messages get delivered to their targets and the data for waiting time distribution fits into a log-normal behaviour of the form
\myeqn{f(w)=\frac{1}{w{\sigma}{\sqrt {2\pi}}}exp(-\frac{({\ln}w-{\mu})^{2}}{2{\sigma}^{2}})}{2(b)}
\eqnreset
As observed in Fig.~\ref{fig:wait}(a), the standard deviation is quite
large as compared to that in Fig.~\ref{fig:wait}(b). This is due to the
fact that for low hub density (0.5$\%$ as in Fig.~\ref{fig:wait}(a))
messages get trapped leading to large fluctuations. The fluctuations are
reduced once the messages get cleared for higher hub density (4.0$\%$ as
in Fig.~\ref{fig:wait}(b)). Thus it is evident that during the congested
phase the waiting time distribution for messages travelling
simultaneously in the network, fits into a Gaussian as compared to the
log-normal fit during the decongested phase. In the decongested phase
messages are not trapped in the lattice and all the messages reach their
respective destinations. This congestion-decongestion transition is
encrypted in the crossover from gaussian to log-normal behaviour in the waiting time distribution for messages travelling in the system. 

To summarize, in this paper, we have studied 
single message and multiple message transport in a gradient network 
based on a substrate lattice of nodes and hubs. The average travel time
on this lattice show $q-$exponential behaviour as a function of hub
density. The power-law tail of this behaviour can be explained in
terms of the log-normal distribution of travel times seen at high hub
densities.    

 Congestion effects are observed when many messages run simultaneously
on the base network. The existence of transport traps can set a limit to
the extent to which congestion is cleared at low hub density. The
waiting time distribution in this regime fits a gaussian. At higher hub
densities  the network is in the decongested phase and the waiting time
distribution shows log-normal behaviour. We note that networks which incorporate geographic clustering and encounter congestion problems arise in many practical situations e.g. cellular networks\cite{Jeong} and air traffic networks \cite{Sinai}. Our results may have relevance in these contexts. 

\begin{acknowledgments}
We wish to acknowledge the support of DST, India under the project SP/S2/HEP/10/2003.
\end{acknowledgments}

\end{document}